\documentclass[12pt]{article}
\begin{document}

\begin{center}
{\huge \bf  Perturbative series and the 1/N expansion for the QED 
$\beta$-function} 
 \\[10mm] 
  S.A. Larin \\ [3mm]
 Institute for Nuclear Research of the
 Russian Academy of Sciences,   \\
 60th October Anniversary Prospect 7a,
 Moscow 117312, Russia
\end{center}

\vspace{30mm}

\begin{abstract}
A comparison of the perturbative series and the 1/N expansion 
for the QED renormalization group
 $\beta$-function in the Minimal Subtraction scheme is performed. The good
agreement between two expansions is found which proves that the 
MS $\beta$-function is under perfect perturbative control.
\end{abstract}

\newpage
1. The nature of perturbative series in Quantum Electrodynamics still remains
an unresolved question, although one can believe that they are asymptotic
sign-alternating series. Then one can hope that the error of the truncated
series of this type is estimated by the value of the first truncated (or
the last included) term of the expansion.
Since QED is the cornerstone of modern Quantum Field
Theory it is rather important to obtain as much  information
as possible concerning its pertubative expansions. 

Quite recently the 5-loop approximation
for the QED renormalization group $\beta$-function in different 
renormalization schemes was 
obtained, first for one active lepton \cite{l} and then for an arbitrary
number $N_F$ of flavors \cite{b,bc}. These results
are obtained after more than twenty years since
the calculation of the 4-loop order \cite{4l}.  

On the other hand there is a calculation
\cite{m} of the first nontrivial leading term 
of the $1/N_F$ expansion for the QED 
$\beta$-function in the MS-scheme. It is quite interesting to compare
the available 5-loop perturbative series and the $1/N_F$ series.
This is the purpose of the present letter.  

2. Let us first cite the $1/N_F$ result for the $\beta$-function
 from the work \cite{m} 
\[
\beta(K)=\frac{2}{3}K+\frac{1}{N_F}\frac{1}{2}K \int_{-K/3}^0 dx
\frac{\Gamma(4+2x)(1+2x)(1+2x/3)(1-x)}{[\Gamma(2+x)]^2\Gamma(3+x)\Gamma(1-x)}
+O(1/N_F^2)
\]
\begin{equation}
\label{bk}
=\frac{2}{3}K+\frac{K^2}{2N_F}\left[1-\frac{11}{2\cdot3}\frac{1}{2}
\left(\frac{K}{3}\right)-\frac{7\cdot11}{2^2\cdot3^2}\frac{1}{3}
\left(\frac{K}{3}\right)^2 \right.
\end{equation}
\[+\left(\frac{107}{2^3\cdot3^2}+2\zeta(3)\right)\frac{1}{4}
\left(\frac{K}{3}\right)^3+\left(\frac{251}{2^4\cdot3^2}-\frac{11}{3}\zeta(3)
+3\zeta(4)\right)\frac{1}{5}\left(\frac{K}{3}\right)^4
\]
\[+\left(\frac{67}{2^5}-\frac{7\cdot11}{2\cdot3^2}\zeta(3)-\frac{11}{2}\zeta(4)
+2\cdot3\zeta(5)\right)\frac{1}{6}\left(\frac{K}{3}\right)^5
\]
\[+\left(\frac{5\cdot7\cdot41}{2^6\cdot3^2}+\frac{107}{2^2\cdot3^2}\zeta(3)
-\frac{7\cdot11}{2^2\cdot3}\zeta(4)-11\zeta(5) \right.
\]
\[\left. \left.
+2\cdot5\zeta(6)-2\zeta^2(3)
\right)\frac{1}{7}\left(\frac{K}{3}\right)^6+...\right]+O(1/N_F^2),
\]
here $K\equiv\alpha N_F/\pi$ is the coupling which has to be held fixed
in the large $N_F$ limit, $\alpha$ being the fine structure constant.

The function  $\beta(K)$  is defined as
\begin{equation}
\label{norm}
 \alpha\beta(K)=\mu\frac{d}{d\mu}\alpha(\mu),
\end{equation}
where $\mu$ is the renormalization scale.

In the numerical form the result of the equation (\ref{bk}) reads
\[
\beta(K)=\frac{2}{3}K+\frac{K^2}{2N_F}\left(1-3.055555556\cdot10^{-1}K
-7.921810700\cdot10^{-2}K^2 \right.
\]
\[
+3.602060109\cdot10^{-2}K^3
+1.438230317\cdot10^{-3}K^4-1.906442773\cdot10^{-3}K^5
\]
\[ \left.
+1.521260392\cdot10^{-4}K^6+3.588903124\cdot10^{-5}K^7+...\right)+O(1/N_F^2).
\]
The radius of convergence of the $\beta(K)$ expansion is $K=15/2$. The authors
of the work \cite{m} checked numerically that the $1/N_F$ term has the only
zero at $K=0$ and is positive in the convergence region. They also found
that for the physical value $N_F=3$ the $1/N_F$ term is never larger than
15\% of the leading term $2K/3$.

Let us now cite the 5-loop result for the $\beta$-function in the MS-scheme
from the work \cite{b}. In the normalization of eq.(\ref{norm}) it is
\begin{equation}
\label{5l}
\beta=8\pi\frac{1}{\alpha} \left\{N_F\left[\frac{4A^2}{3}\right]+4N_FA^3
-A^4\left[2N_F+\frac{44}{9}N_F^2\right] \right.
\end{equation}
\[+A^5\left[-46N_F+\frac{760}{27}N_F^2-\frac{832}{9}\zeta(3)N_F^2
-\frac{1232}{243}N_F^3\right]
\]
\[+A^6\left(N_F\left[\frac{4157}{6}+128\zeta(3)\right]
+N_F^2\left[=\frac{7462}{9}-992\zeta(3)+2720\zeta(5)\right] \right.
\]
\[ \left. \left.+N_F^3\left[-\frac{21758}{81}+\frac{16000}{27}\zeta(3)
-\frac{416}{3}\zeta(4)-\frac{1280}{3}\zeta(5)\right]
+N_F^4\left[\frac{856}{243}+\frac{128}{27}\zeta(3)\right]\right)\right\}
\]
where $A\equiv\frac{\alpha}{4\pi}$.

The numerical form of the above equation for the value $N_F=3$ is
\begin{equation}
\label{num}
\beta=0.63662 \alpha + 0.151982 \alpha^2 - 0.050393 \alpha^3 
- 0.0819407 \alpha^4 + 0.0412278 \alpha^5,
\end{equation}
this is the monotonically increasing function for $\alpha>0$.

We will compare eq.(\ref{bk}) and eq.(\ref{5l}) for 
$N_F=3$.
For $\alpha=1/137$ we have $\beta=0.00465494$ for both equations; for 
$\alpha=0.1$ the result is $\beta=0.0651364$ for eq.(\ref{bk}) 
and $\beta=0.0651236$
for eq.(\ref{5l}); for $\alpha=0.2$ one gets $0.133032$ and $0.132882$
correspondingly; for $\alpha=1$ one gets $0.737883$ for eq.(\ref{bk})
and $0.697496$ for eq.(\ref{5l}).

We see that even for $\alpha=1$ when the convergence of the series (\ref{num})
is quite questionable both results agree within 5\%. Thus two different
expansions (the usual perturbative series and the $1/N_F$ series) give
numerically very close values for the QED $\beta$-function in the wide
interval of $\alpha$. It definitely indicates that both expansions
give good approximations for $\beta(\alpha)$.

  {\bf Acknowledgments.}

The author is grateful to collaborators of the Theory division of INR
for helpful discussions. The work is supported in part by the grant
for the Leading Scientific Schools NS-5590.2012.2.

\end{document}